\documentstyle[multicol,aps,prl,floats]{revtex}

\begin{document}
\draft
\title{
Intermittency of Burgers' Turbulence}
\author{E. Balkovsky$^a$, G. Falkovich$^a$, I. Kolokolov$^b$ 
and V. Lebedev$^{a,c}$}
\address{$^a$ Physics of Complex Systems, Weizmann Institute of Science,
Rehovot 76100, Israel \\ $^b$ Budker Institute of
Nuclear Physics, Novosibirsk 630090, Russia
\\ $^{c}$ Landau Inst. for Theor. Physics, Moscow, Kosygina 2,
117940, Russia}
\date{June 17, 1996}
\maketitle

\begin{abstract}

We consider 
the tails of probability density function (PDF) for
the velocity that satisfies Burgers equation driven by a Gaussian large-scale force.
The saddle-point approximation is employed in the path integral
so that the calculation of the PDF tails
boils down to finding the special field-force configuration (instanton) that 
realizes the extremum of probability. 
For the PDFs of velocity and it's derivatives $u^{(k)}=\partial_x^ku$, the general
formula is found: $\ln{\cal P}(|u^{(k)}|)\propto
-(|u^{(k)}|/{\rm Re}^k)^{3/(k+1)}$.

\end{abstract}
\begin{multicols}{2}

The forced Burgers equation
\begin{equation}
\partial_t u+u\partial_x u 
-\nu\partial_x^2 u=\phi
\label{be1} \end{equation}
describes the evolution of weak one-dimensional acoustic perturbations 
in the reference frame moving with the sound velocity \cite{Burg}. It is 
natural to assume the external force $\phi$ to be $\delta$-correlated 
in time in that frame: $\langle\phi(t_1,x_1)\phi(t_2,x_2)\rangle=
\delta(t_1-t_2)\chi(x_1-x_2)$. 
Then the statistics of $\phi$ is Gaussian and is completely
characterized by $\chi$. We are interested in turbulence excited by
a large-scale pumping with some correlation length $L$ so that $\chi$
does not essentially change at $x\lesssim L$ and goes to zero where $x>L$. 
Besides $L$, the correlation function $\chi$ may be characterized by the parameter 
$\omega=[-(1/2)\chi''(0)]^{1/3}$ having the dimensionality of frequency. 
Then e.g. $\chi(0)\sim L^2\omega^3$. Developed turbulence corresponds 
to a large Reynolds number ${\rm Re}=L^2\omega/\nu\gg1$.

The physical picture of Burgers turbulence is quite clear: arbitrary localized 
perturbation evolves into shock wave with the viscous width of the front,
that gives $k^{-2}$ for the energy spectrum at ${\rm Re}\gg kL\gg1$ 
\cite{Burg,Kra68}. The presence of shocks leads to a strong 
intermittency, PDF of velocity gradients is substantially non-Gaussian 
\cite{GK93} and there is an extreme anomalous scaling for the moments of 
velocity differences $w=u(\rho)-u(-\rho)$:
$\langle w^n\rangle\propto (\rho/L)$ for $n>1$ \cite{Goto94}. 
Simplicity of the equation and transparency of underlying physics make it 
reasonable to hope that consistent formalism for the description of 
intermittency could be developed starting from Burgers equation
\cite{Pol95,BMP95,CY95}.
The most striking manifestation of intermittency is the deviation of 
the PDF tails from Gaussian. 
It has been realized recently  that those tails can be found by 
considering the saddle-point configurations (we call them instantons) in the 
path integral determining the PDF \cite{95FKLM}. 
Note the simultaneous revival of the method of optimal fluctuation
in the description of rare events in disordered metals \cite{KM,FE}.
The instanton formalism has 
been applied to the Burgers equation first by Gurarie and Migdal \cite{95GM} who 
found the right tail $\ln{\cal P}(w)\sim-[w/(\omega\rho)]^3$ determined by 
inviscid behavior of smooth ramps between the shocks. That cubic 
right tail has been earlier predicted by Polyakov from the conjecture on the 
operator product expansion \cite{Pol95}, it corresponds to the same right tail
for the gradients $\ln{\cal P}(u')\sim -(u'/\omega)^3$ 
derived by Gotoh and Kraichnan 
\cite{GK96}. The PDF's
${\cal P}(u')$ and ${\cal P}(w)$ are not symmetric, the asymmetry 
is due to the simple fact that positive gradients are smeared while the 
steepening of negative gradients could be stopped only by viscosity.
Here we describe the viscous instantons that give the left tails of 
PDF's and single-point velocity PDF.

Even though some calculations are lengthy, the simple picture appears as a 
result. Since white forcing pumps velocity by the law $w^2\propto \phi^2t$ 
while the typical time of growth is restricted by the breaking time 
$t\sim L/w$ then the Gaussianity of the forcing 
$\ln{\cal P}(\phi)\propto-\phi^2/\chi(0)$ leads to 
$\ln{\cal P}(w)\sim-[|w|/(L\omega)]^3$. At a shock, 
$w^2\simeq-\nu u'$ so that 
$\ln{\cal P}(u')\sim-[-u'/(\omega{\rm Re})]^{3/2}$.
These simple estimates are confirmed below by consistent calculations.

Following \cite{95FKLM}, we write the high-order moments of the velocity derivatives 
$u^{(k)}=\partial_x^ku$ (including velocity for $k=0$) and 
of the difference $w$ as the path integrals:
\begin{eqnarray} &&
\langle[u^{(k)}]^n\rangle=
\int \!{\cal D}u{\cal D}p\exp\left\{i{\cal I}+
n\ln[u^{(k)}(0,0)]\right\},
\label{si1} \\ &&
\langle w^n\rangle=
\int {\cal D}u{\cal D}p\exp\left\{i{\cal I}+
n\ln[u(0,\rho)-u(0,-\rho)]\right\}.
\label{sio}\end{eqnarray}  
Here, 
${\cal I}=\int dt\,{\cal L}(u,p)$ is the effective
action with the Lagrangian
${\cal L}$ determined by the equation (\ref{be1})
\cite{73MSR,76Dom,76Jan,DP78}:
$${\cal L}=\!\int dx\,(p\partial_t u+pu\partial_xu-
\nu p\partial_x^2u)
+\frac{i}{2}\!\int dx_1dx_2p_1\chi_{12}p_2\,.$$

The main idea implemented here is that the high-order moments 
(for $n\gg1$) are determined by the saddle-point configurations of the 
path integrals (\ref{si1},\ref{sio}). The corresponding saddle-point
equations are
\begin{eqnarray} &&
\partial_t u+u\partial_xu-\nu\partial_x^2u
=-i\int dx'\,\chi(x-x')p(x'),
\label{va2} \\ &&
\partial_t p+u\partial_xp
+\nu\partial_x^2p
=\delta(t)\lambda(x).
\label{vam} \end{eqnarray} 
We should substitute here $
\lambda=[{in}/{a}]\delta'(x)$ with $a(t)=\partial_xu(t,0)$ for gradients,
$\lambda=-({in}/{w})[\delta(x-\rho)-\delta(x+\rho)]$ for differences etc.
A solution of (\ref{va2},\ref{vam}) determines the moments in the
steepest descent approximation:
\begin{equation}
\langle[u^{(k)}]^n\rangle\!\sim\!
[u^{(k)}(0,0)]^n e^{i{\cal I}_{\rm extr}}\!, \,\,\,\,
\langle w^n\rangle \!\sim\![w(0)]^ne^{i{\cal I}_{\rm extr}}.
\label{vaa} \end{equation} 
An instanton solution describes
the field-force configuration that corresponds to the rare fluctuation giving the
main contribution into the high-order moment. This configuration can be also called
optimal fluctuation.

The initial conditions for the instanton equations were formulated
in \cite{95FKLM}: $u\rightarrow0$ at $t\rightarrow-\infty$ and $p=0$ at 
$t>0$.  The role of the last term in (\ref{vam}) is then
reduced to the final condition $p(x)=-\lambda(x)$ 
imposed on $p$ at $t=-0$. Viscosity smears $p$ if to move backwards
in time so that at large negative time both
fields $u$ and $p$ are zero. We will refer to that stage as 
vacuum since it is realized at the absence of the source $\lambda$ 
in the right-hand side of (\ref{vam}).

In this letter, we describe general properties of instantons and find
their dependence on $n$, which gives the
form of PDF tails. Complete analytic descriptions of
the instanton solutions will be published elsewhere.

Since the Lagrangian  
does not explicitly depend on time, 
then the ``energy'' $E$ is conserved by (\ref{va2},\ref{vam}):
$$E=i\int  dx (pu\partial_x u+\nu\partial_x p\partial_x u)
-\frac{1}{2}\int dx_1dx_2p_1\chi_{12}p_2\,.$$
Since ${\cal L}$ does not explicitly depend on
coordinates then the ``momentum'' $J$ is conserved as well:
$iJ=\int dx\,p\partial_xu$.
Because of the conservation laws we should treat 
solutions of (\ref{va2},\ref{vam}) with $E=J=0$ since 
they are zero in the vacuum. That gives the following
saddle-point value of the effective 
action in (\ref{vaa}): ${\cal I}_{\rm extr}=\int dt\,dx\,p\partial_tu$.

Conservation laws help
to understand general properties of the solutions. We consider 
$t=0$, substitu\-te $p(x)=-\lambda(x)$
into $E$ and analyze the balance of different terms.
For the gradient, $E=-na(0)+\omega^3n^2/a^2(0)-n\nu \partial_x^3u(0,0)/a(0)=0$.
Difference between the cases 
of positive and negative $a$
is now clearly 
seen. For $a(0)>0$, 
the viscous contribution to the energy is un\-essential and 
two first terms can compensate each other (see below). 
Contrary, the instanton that gives $a(0)<0$
cannot exist without viscosity.
For the velocity, $J=0$ gives
$\partial_x u(0,0)=0$ which leads to 
$E=\chi(0)n^2/u^2(0)-\nu \partial_x^2u(0,0)=0$. Without viscous term, energy
conservation cannot be satisfied. Let us emphasize that the answer we shall
obtain for the velocity PDF 
does not contain viscosity while it's consistent derivation 
requires the account of the viscous terms in the equations.


Let us first describe the essentially inviscid 
instantons producing the right tails of the PDFs 
for gradients and differences \cite{Pol95,95GM,GK96}. 
At $t=0$, the 
field $p$ is localized near the origin. At moving backwards in time the 
viscosity will spread the field $p$. Nevertheless, 
a positive velocity slope
``compresses'' the field $p$ so that one can expect that the width of $p$ remains
much smaller than $L$.  Then, it is possible to formulate the closed 
system of equations for the quantities $a(t)$ and
$c(t)=-i \int dx\,x\,p(t,x)$ since for narrow $p$ and small $x$ we can put 
$\int dx'\chi(x-x')p(t,x')\to-i\partial_x\chi(x)c(t)\approx2i\omega^3xc(t)$:
\begin{equation}
\partial_tc=2ac, \quad
\partial_ta=-a^2+2\omega^3c.
\label{vca} \end{equation}
The instanton is a separatrix 
solution of (\ref{vca}). 
The initial condition 
$a(0)c(0)=n$ by virtue of the energy conservation 
gives $a(0)=\omega^3c^2(0)/n=\omega n^{1/3}$. 
For differences, $w=2a(0)\rho$.
One can check that 
$ {\cal I}_{\rm extr}=i\int dt\, c\partial_t a\sim a(0)c(0)=n $ which 
is negligible in comparison with $n\ln[a(0)]$ so that
$\langle(u')^n\rangle\sim[a(0)]^n \sim \omega^n n^{n/3}$ 
which gives the right cubic tails of the PDFs 
$\ln{\cal P}(u')\sim-(u'/\omega)^3$ \cite{GK96} and
$\ln{\cal P}(w)\sim-[w/(\rho\omega)]^3$ 
\cite{Pol95,95GM}.  
One can show that the width of $p$ is much less than $L$ through the 
time of evolution $T\sim n^{-1/3}\omega^{-1}$ 
giving the main contribution into the action \cite{95GM}. 
The right tails of ${\cal P}(u')$ and ${\cal P}(w)$ are thus
universal i.e. independent of the large-scale properties of the pumping. 
Above consideration does not imply that the instanton is completely inviscid,
it may well have viscous shock at $x\sim L$, this has no influence
on the instanton answer (since $p$ is narrow) while may influence the fluctuation
contribution i.e. predexponent in the PDF.


The main subject of this paper is the analysis of the instantons that give
the tails of ${\cal P}(u)$ and the left tails of ${\cal P}(u')$ 
and ${\cal P}(w)$ corresponding to negative $a$, $w$. 
Even though the field $p$ is narrow at $t=0$, we cannot use the simple 
system (\ref{vca}) to describe those instantons. 
The reason is that sweeping by a negative velocity slope provides for 
stretching (rather than compression) of the field $p$ at moving backwards 
in time. As a result, the support of $p(x)$ stretches up to $L$ so that
one has to account for the given form of the pumping correlation function
$\chi(x)$ at $x\simeq L$. This leads to a nonuniversality of ${\cal P}(u)$ and of 
the left tails of ${\cal P}(u')$ and ${\cal P}(w)$ which depend
on the large-scale properties of the pumping. As we shall see, the form of
the tails is universal, nonuniversality is related to a single constant in PDF.
Additional complication in analytical description is due to
the shock forming from negative slope near the
origin. The shock cannot be described in terms of the inviscid 
equations so that we should use the complete system
(\ref{va2},\ref{vam}) to describe what can be called
viscous instantons.

Apart from a narrow front near $x=0$, the velocity field 
has $L$ as the only characteristic scale of change. The life time $T$ of
the instanton is then determined by the moment when the position of $p$ 
maximum reaches $L$ due to sweeping by the velocity 
$u_0$: $T\sim L/u_0$. Such a velocity $u_0$ itself has been created during 
the time $T$ by the forcing so that $u_0\sim|c|_{max}TL\omega^3$. 
To estimate the maximal value of $|c(t)|$, let us consider the backward 
evolution from $t=0$. We first notice that the width of $p$ (which was zero 
at $t=0$) is getting larger than the width of the velocity front $\simeq u_0/a$
already after the short time $\simeq a^{-1}$. After that time, the 
values of $c$ and $a$ are of order of their values at $t=0$. 
Then, one may consider that $p(t,x)$ propagates (backwards in time) 
in the almost homogeneous velocity field $u_0$ so that 
$$\partial_t c=-i\int_{-\infty}^{\infty} dx\, xup_x\approx 2iu_0\int_0^\infty dx\, p
\ .$$ The (approximate) integral
of motion $i\int dx\, p$ can be estimated by it's value at $t=0$ which 
is $n/2u_0$. Therefore, we get $c_{max}\simeq nT$ so that
$T\simeq n^{-1/3}\omega^{-1}$ and $u_0\simeq L\omega n^{1/3}$. 
At the viscosity-balanced shock, the velocity $u_0$ and the gradient $a$ 
are related by $u_0^2\simeq\nu a$ so that $a(0)\simeq \omega{\rm Re}\,n^{2/3}$.

Let us briefly describe now the consistent analytic procedure of the derivation of
the function $c(t)$ that confirms above estimates. We use the
Cole-Hopf substitution \cite{Burg} for the velocity $\partial_x\Psi=-{u}\Psi/{2\nu}$
and introduce $P=2\nu\partial_xp/\Psi$.
The saddle-point equations for $\Psi$ and $P$ 
\begin{eqnarray} &&
\partial_t\Psi-\nu\partial_x^2\Psi+\nu F\Psi=0,
\label{ha7} \\ &&
\partial_t P+\nu\partial_x^2P-\nu FP
-{2\nu}\lambda'(x)\delta(t)\Psi^{-1}=0
\label{ha3} \end{eqnarray}
contain $F$ determined by $\partial_xF(t,x)=-{i}
\int dx'\chi(x-x')p(t,x')/{2\nu^2}$ and fixed by the condition $F(t,0)=0$. 
We introduce the evolution operator $\hat U(t)$
which satisfies the equation $\partial_t\hat U=\hat H\hat U$ with
$\hat H(t)=\nu(\partial_x^2-F)$. It is remarkable that one 
can develop the closed description
in terms of two operators $\hat A=\hat U^{-1} x\hat U$ and $\hat B=
\hat U^{-1}\partial_x\hat U$:
$$\partial_t\hat A=-2\nu\hat B\,,\quad \partial_t\hat B=-\nu F_x(t,\hat A)\ .$$
Since we are 
interesting in the time interval when $p(t,x)$ is narrow, it is enough for our
purpose to consider $x\ll L$ where $F(t,x)=c(t)x^2\omega^3/2\nu^2$. Further
simplification can be achieved in this case and the closed ODE for $c(t)$ can
be derived after some manipulations:
$$(\partial_t c)^2=4\omega^3c^3+16\xi_2^2+4\omega^3\xi_1^3\ ,$$
where $\xi_1\!=i\int\! dx\lambda(x) x$ and $4\xi_2\!=-{i}\int\! dx\lambda(x)
\partial_x[xu(0,x)]$. Integrating we get
\begin{eqnarray}&&
t=\frac{1}{2}
\int_{c(0)}^{c}\frac{dx}{\sqrt{\omega^3x^3+4\xi_2^2+\xi_1^3}}\ ,
\label{b7}\end{eqnarray}
which describes $c(t)$ in an implicit form. Further analysis depends on the
case considered. For the gradients, we substitute $\xi_1=n/a_0$ and $\xi_2=-n/2$
and see that, as time goes backwards, negative $c(t)$ initially decreases by 
the law $c(t)=c(0)+2nt$ until $T= \omega^{-1}(n/2)^{-1/3}$ then it grows
and the approximation looses validity when $c(t)$ approaches zero and the account
of the pumping form $\chi(x)$ at $x\simeq L$ is necessary. Requiring the width
of $p(x)$ at this time to be of order $L$ we 
get the estimate $a(0)\simeq \omega{\rm Re}\,n^{2/3}$ and thus confirm the above
picture.
The main contribution to the saddle-point value (\ref{vaa})
is again provided by the term $[\partial_xu(0,0)]^n$ 
and we find $\langle(u')^n\rangle\simeq[a(0)]^n\simeq
(\omega{\rm Re})^n n^{2n/3}$, which corresponds to the following left 
tail of PDF at $u'\gg \omega{\rm Re}$
\begin{equation} 
{\cal P}(u')\propto 
\exp[-C(-u'/\omega{\rm Re})^{3/2}]\ .
\label{an2} \end{equation}
For higher derivatives $u^{(k)}$, by using (\ref{b7}) we get initial growth 
$c(t)=c(0)+n(k+1)t$ which gives
$u^{(k)}(0,0)\sim N^{k+1}L^{1-k}\omega {\rm Re}^{k}$ leading to
$\langle[u^{(k)}]^n\rangle\sim\omega {\rm Re}^{k} 
L^{1-k} n^{(k+1)/3}$ which can be rewritten in terms of PDF: 
\begin{eqnarray} &&
{\cal P}\left(|u^{(k)}|\right)\!\propto\!
\exp\left[-C_k\left({|u^{(k)}|L^{k-1}}/
{\omega{\rm Re}^k}\right)^{3/(k+1)}\right].
\label{hd4}\end{eqnarray}
Note that the non-Gaussianity increases with increasing $k$. On the other hand,
the higher $k$ the more distant is the validity region of (\ref{hd4}): 
$u^{(k)}\gg u^{(k)}_{\rm rms}\sim L^{1-k}\omega{\rm Re}^k$.

For the differences, $\xi_1={2n\rho_0}/{w}$ and
$4\xi_2=-{n}[1+{2\rho_0u_x(0,\rho_0)}/{w}]$ and we get 
$\langle w^n\rangle\simeq (L\omega)^nn^{n/3}$
which corresponds to the cubic left tail 
\begin{equation}  {\cal P}(w)\propto \exp\{-B[w/(L\omega)]^3\}
\label{an3} \end{equation}
valid at $w\gg L\omega$. In the intermediate region $L\omega\gg w\gg\rho\omega$,
there should be a power asymptotics which is the subject of current debate
\cite{Pol95,GK96,KS96}.
It is natural that $\rho$-dependence of ${\cal P}(w)$ cannot be found in a 
saddle-point approximation; as a predexponent, it can be obtained only at the
next step by calculating the contribution of fluctuations around the instanton
solution. This is consistent with the known fact that the scaling exponent
is $n$-independent for $n>1$: $\langle w^n(\rho)\rangle\propto\rho$.

For the velocity, $\lambda(x)=-{in}\delta(x)/u(0,0)$ is an even function
so that  $F$ is a linear (rather than quadratic)
function of $x$ for narrow $p$: $F(x)={\chi(0)bx}/{2\nu^2}$ with
$b=-{i}\int dx p(x)$. Direct calculation shows that energy and momentum
conservation makes $b$ time independent: $b=n/u(0,0)$.
It is easy then to get the $n$-dependence of $u(0,0)$: 
Velocity stretches the field $p$ so that the width of $p$
reaches $L$ at $T\simeq L/u(0,0)$ while the velocity itself is produced by
the pumping during the same time: $u(0,0)\simeq \chi(0)bT=\chi(0)nT/2u(0,0)
\simeq n\chi(0)L/u(0,0)$. That gives $u(0,0)\simeq L\omega n^{1/3}$ and
$$ {\cal P}(u)\propto \exp\{-D[u/(L\omega)]^3\}\ .$$ 
The product $L\omega$ plays the role of the root-mean square velocity
$u_{\rm rms}$. The numerical factors $C$, $B$ and $D$ 
are determined by the evolution at $t\simeq T$ i.e. by the behavior of pumping
correlation function $\chi(x)$ at $x\simeq L$.

We thus found the main exponential factors in the PDF tails. 
Complete description of the tails requires the analysis
of the fluctuations around the instanton which will be the subject of
future detailed publications. Here, we briefly outline some important
steps of this analysis. The account of the fluctuations in the Gaussian
approximation is straightforward and leads to the shift of ${\cal I}_{\rm extr}$
insignificant at $n\gg1$. However, the terms of the perturbation theory with 
respect to the interaction of fluctuations are infrared divergent (proportional
to the total observation time). That means that there is a soft mode which
is to be taken into account exactly. Such an approach has been already developed
in \cite{95FKLM} for the simpler problem of the PDF tails for a passive scalar
advected by a large-scale velocity where the comparison with the exactly
solvable limits was possible. 
A soft mode usually corresponds to a global symmetry
with a continuous group: if one allows the slow spatio-temporal
variations of the parameters of the transformation then small variations of
the action appears. 
Our instantons break Galilean invariance so that the respective
Goldstone mode has to be taken into account. Namely, under the transformation
\begin{equation}x\rightarrow x-r,\ u(x)\rightarrow u(x-r)+v,\ 
r=\int_t^0v(\tau)d\tau\ ,\label{sym}\end{equation}
the action is transformed as ${\cal I}\rightarrow{\cal I}-i\int dxdtp\partial_tv$.
The source term $\int dxdt\lambda u$ is invariant with respect to (\ref{sym}) for
antisymmetric $\lambda(x)$. To integrate exactly along the direction specified by
(\ref{sym}) in the functional space we use Faddeev-Popov trick 
inserting  
the additional factor 
\begin{equation}
1=\int{\cal D}v(t)\delta\left[u\biggl(t,\int_t^0v(\tau)d\tau\biggr)-
v(t)\right] {\cal J}\ .
\label{unity}\end{equation}
into the integrand in (\ref{si1},\ref{sio}).
Jacobian ${\cal J}$ is determined by a regularization of (\ref{sym})
according to our choice of the retarded regularization for the initial
integral: at 
discretizing time we put $\partial_tu+u\partial_xu
\rightarrow ({u_n-u_{n-1}})/{\epsilon}+u_{n-1} u'_{n-1}$ (otherwise, 
some additional $u$-dependent term appears \cite{DP78}). 
The discrete version of (\ref{sym}),
$p_n(x)\rightarrow p_n(x-\epsilon\sum_{j=n}^{N-1}v_j)$,
$u_n(x)\rightarrow u_n(x-\epsilon\sum_{j=n}^{N-1}v_j)-v_n$, $u_N(x)\rightarrow
u_N(x)-v_N$ gives 
$${\cal J}=\exp\left[\int_{-T}^0dtu'\biggl(t,\int_t^0 v(\tau)d\tau\biggr)\right]\ .$$
Substituting (\ref{unity}) into (\ref{si1},\ref{sio})
and making (\ref{sym}) we calculate $\int{\cal D}v$ as a Fourier integral 
(the saddle-point method is evidently
inapplicable to such an integration) and conclude that after the integration
over the mode (\ref{sym}) the measure ${\cal D}u{\cal D}pe^{i{\cal I}}$
acquires the additional factor
$$\prod\limits_t\!\delta\biggl[\int\partial_t^2 p(t,x)dx\biggr]
\delta[u(t,0)]\exp\left[
\int_{-T}^0\!u'(t,0)dt\right].$$
The last (jacobian) term here exactly corresponds
to the term $u'{\cal P}(u')$  in the equation for ${\cal P}(u')$ 
derived in \cite{GK93,GK96}. This term 
makes the perturbation theory for the fluctuations around the instanton to be
free from infrared 
divergences, the details will be published elsewhere.

Let us summarize.
At smooth almost inviscid ramps, velocity differences and gradients are 
positive and linearly related $w(\rho)\approx 
2\rho u'$ so that the right tails of PDFs have the same cubic form 
\cite{Pol95,95GM,GK96}.
Those tails are universal i.e. they are determined by a single characteristics
of the pumping correlation function $\chi(r)$, namely, by it's second
derivative at zero $\omega=[-(1/2)\chi''(0)]^{1/3}$. 
Contrary, the left tails found here contain nonuniversal constant which 
depends on a large-scale behavior of the pumping. The left tails
come from shock fronts where $w^2\simeq -\nu u'$ so that cubic tail 
for velocity differences (\ref{an3}) corresponds to semi-cubic tail for gradients
(\ref{an2}).
The formula (\ref{an3}) is valid for $w\gg u_{\rm rms}\simeq L\omega$
where ${\cal P}(w)$ should coincide with 
a single-point ${\cal P}(u)$ since the  
probability is small for both $u(\rho)$ and $u(-\rho)$ being large 
simultaneously. Indeed, we saw that the tails of
$\ln{\cal P}(u)$ at $u\gg u_{\rm rms}$ are cubic as well.
Note that 
(\ref{hd4}) is the same as obtained for decaying turbulence with white (in space)
initial conditions
by a similar method employing the saddle-point approximation in the 
path integral with time as large parameter \cite{Avel}. 
That, probably, means that white-in-time forcing
corresponds to white-in-space initial conditions. Note that
if the pumping has a finite correlation time $\tau$ then our
results, strictly speaking, are valid for 
$u,w\ll L/\tau$ and $u'\ll1/\tau$.

We are grateful to M. Chertkov, V. Gurarie, D. Khmelnitskii, R. Kraichnan 
and A. Polyakov for useful discussions. 
This work was supported 
by the Minerva Center for Nonlinear Physics (I. K. and V. L.), 
by the Minerva Einstein Center (V. L.), by the Israel Science Foundation
founded by the Israel Academy (E.B.) and by 
the Cemach and Anna Oiserman Research Fund (G.F.).

\end{multicols}

\end{document}